\documentstyle[12pt]{article}

\newcommand{\lab}[1]{\label{#1}}

\newcommand{\rf}[1]{\ref{#1}}

\newcommand{\wt}{\widetilde}

\newcommand{\spc}{{\hspace{1em}}}

\newcommand{\filename}{\underline{\Large Filename : \tt\jobname.tex}}

\newcommand{\rap}[2]
{\setbox1=\hbox{#1}%
\setbox2=\hbox to\wd1{\hss #2\hss}%
\mbox{\rlap{\box1}\box2}}

\newcommand{\J}[4]{{\sl #1} {\bf #2} (#3) #4}
\newcommand{\NP}{Nucl.\ Phys.}
\newcommand{\PL}{Phys.\ Lett.}

\begin{document}

\baselineskip=18pt plus 0.2pt minus 0.1pt

\begin{titlepage}
\title{
\hfill\parbox{4cm}
{\normalsize KUNS-1457\\HE(TH)~97/11\\{\tt hep-th/9706144}}\\
\vspace{1cm}
Born-Infeld Action and Chern-Simons Term\\
from Kaluza-Klein Monopole in M-theory
}
\author{
Yosuke Imamura\thanks{{\tt imamura@gauge.scphys.kyoto-u.ac.jp}}
{}\thanks{
Supported in part by Grant-in-Aid for Scientific
Research from Ministry of Education, Science and Culture
(\#5416).}
\\[7pt]
{\it Department of Physics, Kyoto University, Kyoto 606-01, Japan}
}
\date{}

\maketitle
\thispagestyle{empty}

\begin{abstract}
\normalsize
We investigate the zero modes
of the Kaluza-Klein monopole in M-theory
and show that the Born-Infeld action
and the Chern-Simons term
of a D6-brane are reproduced
to quadratic order in the field strength
of the $U(1)$ field on the brane.
\end{abstract}

\end{titlepage}

%%%%%%%%%%%%%%%%%%%%%%%%%%%%%%%%%%%%%%%%%%%%%%%%%%%%%%%%%%%%%%%%%%%%%%%%%%%
%%%%%%%%%%%%%%%%%%%%%%%%%%%%%%%%%%%%%%%%%%%%%%%%%%%%%%
\section{Introduction}
M-theory \cite{WittenM} is a powerful tool for understanding
various results in string theory.
Although only its low energy effective theory is known,
if we believe it, we can derive quantities in string theory
without carrying out stringy calculations \cite{PowerM,Alwis,Alwis2}.

In string theory, it is known that the D$p$-brane is governed by the
following action,
which is the sum of the Born-Infeld action and the Chern-Simons action
(in present paper, we ignore fermion terms):
\begin{eqnarray}
S_{{\rm D}p}&=&S_{\rm BI}+S_{\rm CS},\\
S_{\rm BI}&=&\frac{e^{-\phi_0}}{(2\pi)^pl_s^{p+1}}\int d^{p+1}\sigma
\,{\rm det}^{\frac{1}{2}}\left[-G_{ij}+(2\pi l_s^2)V_{ij}\right]
    \lab{BIaction},\\
S_{\rm CS}&=&\sqrt{2\pi}(2\pi l_s)^{3-p}\int_{p+1}Ce^{(2\pi l_s^2)V},
\lab{CSaction}
\end{eqnarray}
\begin{equation}
V_{ij}=\nabla_iV_j-\nabla_jV_i-\frac{1}{2\pi l_s^2}B_{ij},
\end{equation}
where $G_{ij}$ and $B_{ij}$ are the pull-backs to the D-brane world volume
of the spacetime metric $G_{MN}$ and the NS-NS two form field $B_{MN}$.
$C$ is the sum of the R-R 1, 3, 5 and 7-form fields.
$V=(1/2)V_{ij}d\sigma^i\wedge d\sigma^j$ is the field strength
 of the $U(1)$ gauge field $V_i$
living on the D$p$-brane shifted by $B_{ij}$.
$V_i$ is normalized so that
 the charge of the string endpoint is one.
$l_s=\sqrt{\alpha'}$ is the string length scale.

For $p=2$ and $4$, $S_{{\rm D}p}$ can be derived
from the actions of M2 and M5-brane
in M-theory \cite{Alwis2}, and for $p=0$,
it is obtained as the effective action of the Kaluza-Klein mode of
the gravity multiplet in eleven dimensional supergravity.
For $p=6$, it has been expected that the M-theory configuration
 corresponding to D6-brane is a Kaluza-Klein monopole
 \cite{Townsend}.
Kaluza-Klein monopole
is a Dirac monopole of $U(1)$ gauge field which appears
when a higher dimensional theory is compactified.
{}From the viewpoint of the higher dimensional theory
the monopole is a gravitational instanton.

The purpose of this paper is to show  that
the Born-Infeld action (\rf{BIaction}) and
the Chern-Simons one (\rf{CSaction})
on a D6-brane are reproduced
as the effective action of
the zero modes on the Kaluza-Klein monopole.
Let us take the static gauge $\sigma^i=y^i$.
Then (\rf{BIaction}) and (\rf{CSaction}) are
expanded as follows
(we put $p=6$ and neglect higher order terms of the Born-Infeld action),
\begin{eqnarray}
S_{\rm BI}&=&-\frac{e^{-\phi_0}}{(2\pi)^6l_s^7}\int d^7y
\left[\frac{1}{2}(\partial_i\phi^a)^2
       +\frac{1}{4(2\pi l_s^2)^2}V_{ij}^2\right],
      \lab{BIexpand}\\
S_{\rm CS}&=&\frac{\sqrt{2\pi}}{(2\pi l_s)^3}
              \int d^7y\,\epsilon^{i_1\cdots i_7}
              \left[\frac{1}{7!}C_{i_1\cdots i_7}
                   +\frac{(2\pi l_s^2)}{5!\cdot2}
                                C_{i_1\cdots i_5}V_{i_6i_7}
                   \right.\nonumber\\&&\hspace{2em}\left.
                   +\frac{(2\pi l_s^2)^2}{3!\cdot2\cdot2\cdot2}
                                C_{i_1i_2i_3}
                                V_{i_4i_5}V_{i_6i_7}
                   +\frac{(2\pi l_s^2)^3}{3!\cdot2\cdot2\cdot2}C_{i_1}
                                V_{i_2i_3}V_{i_4i_5}V_{i_6i_7}\right].
\lab{CSexpand}
\end{eqnarray}
We shall show that the $V^2$ term in (\rf{BIexpand}) and the 
first three terms in (\rf{CSexpand}) are correctly reproduced
from the Kaluza-Klein monopole.

Our conventions are similar to the ones in \cite{Alwis}.
Instead of explaining them, we present the actions
of the eleven and ten dimensional theories:
\begin{itemize}
\item M-theory %%%%%%%%%%%%%%%%%%%%%%%%%%%%%%%%%%%%%%%

\begin{eqnarray}
S_{M\rm-kin}&=&-\int d^{11}x
    \left[\frac{1}{2\kappa_{11}^2}\sqrt{-g}R
         +\frac{1}{2\cdot4!}\sqrt{-g}K_{KLMN}K^{KLMN}
    \right],\lab{Maction}\\
S_{M\rm-CS}&=&
         -\frac{\sqrt{2\kappa_{11}^2}}{3!3!4!4!}
       \int d^{11}x
   \epsilon^{ABCDEFGHIJK}
   A_{ABC}K_{DEFG}K_{HIJK},
   \lab{MCS}\\
S_{M2}&=&\frac{g_{M2}}{3!}
        \int_{M2}dx^L\wedge dx^M\wedge dx^N A_{LMN},\lab{SM2}\\
S_{M5}&=&\frac{g_{M5}}{5!}\int_{M5}dx^{N_1}\wedge\cdots\wedge dx^{N_6}
                         A_{N_1\cdots N_6},\lab{SM5}
\end{eqnarray}
\begin{equation}
K_{KLMN}=\nabla_KA_{LMN}-\nabla_LA_{MNK}+\nabla_MA_{NKL}-\nabla_NA_{KLM},
\end{equation}
\begin{equation}
2\kappa_{11}^2=(2\pi)^8l_s^9,\spc
g_{M2}=(2\pi)^2l_s^{\frac{3}{2}},\spc
g_{M5}=(2\pi)^{-1}l_s^{-\frac{3}{2}}.
\end{equation}

\item Type IIA theory %%%%%%%%%%%%%%%%%%%%%%%%%%%

\begin{eqnarray}
S_{NS-NS}&=&-\int d^{10}x\sqrt{-g}\frac{1}{2\kappa_{10}^2}\left[R
           -4(\partial_M\phi)^2
           +\frac{1}{2\cdot3!}(H_{LMN})^2\right],\lab{NS-NSaction}\\
S_{R-R}&=&-\int d^{10}x\sqrt{-g}\left[\frac{1}{4}(F_{MN})^2
         +\frac{1}{2\cdot4!}(F_{KLMN})^2\right],
\lab{RRaction}\\
H_{LMN}&=&\nabla_LB_{MN}+\nabla_MB_{NL}+\nabla_NB_{LM},\\
F_{MN}&=&\nabla_MC_N-\nabla_NC_M,\\
F_{KLMN}&=&\nabla_KC_{LMN}-\nabla_LC_{MNK}+\nabla_MC_{NKL}-\nabla_NC_{KLM},
\end{eqnarray}
\begin{equation}
2\kappa_{10}^2=(2\pi)^7l_s^8e^{2\phi}.
\end{equation}
\end{itemize}

%%%%%%%%%%%%%%%%%%%%%%%%%%%%%%%%%%%%%%%%%%%%%%%%%%%%%%%%%%%%%%%%%%%%%%%

%%%%%%%%%%%%%%%%%%%%%%%%%%%%%%%%%%%%%%%%%%%%%%%%%%%%%%
\section{Taub-NUT manifold}
Taub-NUT manifold is a kind of four dimensional gravitational instanton.
Far from the center of the instanton, the structure of the manifold is
roughly $R^3\times S^1$.
The $S^1$ bundle over $S^2\subset R^3$ enclosing the instanton
has Pontryagin index one and is topologically equivalent to $S^3$.
Taking the spherical coordinate $(r,\theta,\phi)$ for $R^3$
and parameterizing the fiber $S^1$ by $\psi$,
the metric of the manifold reads
\begin{equation}
ds_{\rm Taub-NUT}^2=\frac{r+m}{r-m}dr^2
    +(r^2-m^2)(d\theta^2+\sin^2\theta d\phi^2)
    +(4m)^2\frac{r-m}{r+m}\left(d\psi+\frac{1}{2}\cos\theta d\phi\right)^2.
\lab{TaubNut}
\end{equation}
This metric has one scale parameter $m$,
representing the size of $S^1$ fiber at $r\rightarrow\infty$.
The ranges of the coordinates are
\begin{equation}
m\leq r,\spc
0\leq\theta\leq\pi,\spc
0\leq\phi<2\pi,\spc
0\leq\psi<2\pi.
\end{equation}
The center of the instanton corresponds to $r=m$.
This manifold is homeomorphic to $R^4$.

Taub-NUT manifold is not only Ricci flat but also hyper K\"ahler,
so that its holonomy group is $SU(2)_L\subset SU(2)_L\times SU(2)_R$.
Therefore, half of the (global) $32$ supersymmetries
corresponding to left handed spinor on the manifold is broken,
while the unbroken $16$ supersymmetries corresponds
to supersymmetries on the D6-brane.

%%%%%%%%%%%%%%%%%%%%%%%%%%%%%%%%%%%%%%%%%%%%%%%%%%%%%%%%%%%%%%

Let us construct a Kaluza-Klein monopole solution corresponding to a
D6-brane in type IIA theory with string coupling $e^{\phi_0}$.
The boundary conditions of type IIA supergravity are
\begin{equation}
\phi(r\rightarrow\infty)=\phi_0,\spc
g_{\mu\nu}^{(10)}(r\rightarrow\infty)=\eta_{\mu\nu}.
\lab{IIAbc}
\end{equation}
Since the metrics in ten and eleven dimensions are related by
\begin{equation}
g_{\mu\nu}^{(11)}=e^{-\frac{2}{3}}g_{\mu\nu}^{(10)},\spc
g_{11,11}^{(11)}=l_s^2e^{\frac{4}{3}\phi},\spc
0\leq x^{11}<2\pi,\lab{rescale}
\end{equation}
(\rf{IIAbc}) implies the following boundary conditions for the eleven
dimensional metric:
\begin{eqnarray}
g_{\mu\nu}^{(11)}(r\rightarrow\infty)&=&e^{-\frac{2}{3}\phi_0}\eta_{\mu\nu},
\lab{GmnCond}\\
g_{11,11}^{(11)}(r\rightarrow\infty)&=&l_s^2e^{\frac{4}{3}\phi_0}.
\lab{G1111Cond}
\end{eqnarray}
Using the Taub-NUT solution (\rf{TaubNut}),
the metric satisfying the condition (\rf{GmnCond}) is
immediately given as
\begin{equation}
ds^2=e^{-\frac{2}{3}\phi_0}
\left[\eta_{ij}dy^idy^j+ds_{\rm Taub-NUT}^2\right],
\lab{11dimmetric}
\end{equation}
where $y^i$ $(i=0,\ldots,6)$ is the flat coordinate
on $R^7$ parallel to the D6-brane.
{}From the second condition (\rf{G1111Cond}),
the scale parameter $m$ is fixed as
\begin{equation}
4m=l_se^{\phi_0}.
\end{equation}

%%%%%%%%%%%%%%%%%%%%%%%%%%%%%%%%%%%%%%%%%%%%%%%%%%%%%%
%%%%%%%%%%%%%%%%%%%%%%%%%%%%%%%%%%%%%%%%%%%%%%%%%%%%%%%%%%%%%%%%%%%%%%%%%%%
\section{$U(1)$ field on Kaluza-Klein monopole}\lab{sec:FB}
Dividing the eleven dimensional coordinates $x^M$ ($M=0,\ldots,9,11$)
into the seven dimensional flat coordinates $y^i$ ($i=0,\ldots,6$) and
the coordinates on Taub-NUT manifold $x^\mu$
($x^7\equiv r$, $x^8\equiv\theta$, $x^9\equiv\phi$ and $x^{11}\equiv\psi$),
the three form field $A_{LMN}$ is classified into four types of fields:
\begin{displaymath}
A_{LMN}\spc\rightarrow
\spc A_{\lambda\mu\nu},
\spc A_{i\mu\nu},
\spc A_{ij\mu},
\spc A_{ijk}.
\end{displaymath}
Our goal in this section is to derive the $(B-F)^2$-term
and hence we are interested in $A_{i\mu\nu}$ and $A_{ij\mu}$.
Since we would like to study the massless modes,
we should express $A_{ij\mu}$ and $A_{i\mu\nu}$
as products of zero modes on Taub-NUT manifold
and functions depending only on the flat coordinate $y^i$:
\begin{eqnarray}
A_{ij\mu}(x^\mu,y^i)&=&A_\mu(x^\mu)B_{ij}(y^i),
\lab{bunri1}\\
A_{i\mu\nu}(x^\mu,y^i)&=&A_{\mu\nu}(x^\mu)V_i(y^i).
\lab{bunri2}
\end{eqnarray}
$V_i$ becomes $U(1)$ field on the brane and $B_{ij}$ plays the role of
NS-NS two form field in the bulk.

Taub-NUT manifold has $SU(2)\times U(1)$ isometry,
where $SU(2)$ represents a rotation of the base manifold $R^3$
and $U(1)$ corresponds to a global shift along the $S^1$ fiber.
Therefore, we can classify modes on the manifold
by two integers $L$ and $N$%
\footnote{To be precise, when $N$ is odd, $L$ must be an half integer
because of the anti-periodic boundary condition
due to Dirac strings with $1/2$ flux at $\theta=0$ and $\pi$.}.
$L$ is the angular momentum with respect to the $SU(2)$ rotation
and $N$ labels the Kaluza-Klein momentum along the $S^1$ fiber.
In the type IIA theory context, $N$ represents the D-particle charge,
and we are interested in the modes with $N=0$ in this paper.
The vector field $V_i$ and the pull-back $B_{ij}$ of the two form field
are scalars with respect to the $SU(2)$ rotation,
so that we should consider the modes with $(L,N)=(0,0)$.
The $(0,0)$ modes of the vector and the two form fields
on Taub-NUT manifold are expressed as follows:
\begin{eqnarray}
A_{\mu}&=&\left(f_1(r),0,\frac{1}{2}f_2(r)\cos\theta,f_2(r)\right),
       \lab{AmExpand}\\
A_{\mu\nu}&=&\left(\begin{array}{cccc}
0 & 0 & \frac{1}{2}g_1(r)\cos\theta & g_1(r) \\
0 & 0 & g_2(r)\sin\theta & 0 \\
- -\frac{1}{2}g_1(r)\cos\theta & -g_2(r)\sin\theta & 0 & 0 \\
- -g_1(r) & 0 & 0 & 0
\end{array}\right),
\lab{AmnExpand}
\end{eqnarray}
where $f_1$, $f_2$, $g_1$ and $g_2$ are functions depending only on $r$.

The vector field zero mode is defined by following equation of motion:
\begin{equation}
\nabla^\alpha\nabla_{[\alpha}A_{\mu]}=0.
\lab{AmEq}
\end{equation}
Note that this equation is not equivalent
to $\nabla_{[\alpha}A_{\mu]}=0$ since $A_\mu$ needs not vanish at
infinity $r\rightarrow\infty$.
Substituting (\rf{AmExpand}) into (\rf{AmEq}), we get
\begin{equation}
- -4m^2f_2+2r(r^2-m^2)f'_2+(r^2-m^2)^2f''_2=0,
\end{equation}
which is easily solved to give
\begin{equation}
f_2=c_1\frac{r+m}{r-m}+c_2\frac{r-m}{r+m}.
\lab{A4is}
\end{equation}
The function $f_1$ does not appear in the equation of motion.
This is a reflection of the fact that $f_1$ is a gauge degree of freedom.
Because the first term in (\rf{A4is}) is singular at the center $r=m$,
$c_1$ should be zero.
The normalization factor $c_2$ should be determined
so that the kinetic term of $B_{ij}$
in the asymptotic region agrees with the third term in (\rf{NS-NSaction}).
Substituting (\rf{A4is}) into the M-theory action (\rf{Maction}),
we get
\begin{eqnarray}
- -\frac{1}{2\cdot3!}\int d^{11}x\sqrt g K_{\mu ijk}K^{\mu ijk}
&=&-\frac{1}{2\cdot3!}\int dx^{11}\sqrt{g_{11,11}}\times
                   A_\mu A^\mu\times
                   \int d^{10}x\sqrt{g}H_{ijk}H^{ijk}\nonumber\\
&=&-\frac{1}{2\cdot3!}2\pi l_se^{\frac{2}{3}\phi_0}\times
                     \frac{c_2^2}{l_s^2}e^{-\frac{4}{3}\phi_0}\times
                   \int d^{10}x\sqrt{g}H_{ijk}H^{ijk},
\end{eqnarray}
which, after rescaling the metric according to (\rf{rescale}),
is reduced to
\begin{equation}
- -\frac{1}{2\cdot3!}\frac{2\pi}{l_s}e^{-2\phi_0}c_2^2
                   \int d^{10}x\sqrt{g}H_{ijk}H^{ijk}.
\end{equation}
Comparing this equation to (\rf{NS-NSaction}),
we can fix the constant $c_2$:
\begin{equation}
c_2=-\frac{1}{(2\pi)^4l_s^\frac{7}{2}}.
\end{equation}

Next, let us construct the zero mode of the two form $A_{\mu\nu}$.
It should satisfy the equation of motion
 $\nabla^\alpha\nabla_{[\alpha}A_{\mu\nu]}=0$,
which is equivalent to
\begin{equation}
\nabla_{[\lambda}A_{\mu\nu]}=0.
\lab{AmnEq}
\end{equation}
Because the holonomy group of Taub-NUT manifold is $SU(2)$,
the zero mode of the self-dual part of $A_{\mu\nu}$, which is singlet
with respect to the $SU(2)$, is a covariantly constant tensor
and is not normalizable.
To get a normalizable solution, we impose the anti-self-duality
condition on $A_{\mu\nu}$:
\begin{equation}
A_{\mu\nu}=-\sqrt{-g}\,\epsilon_{\mu\nu\rho\sigma}A^{\rho\sigma}.
\lab{anti self dual cond}
\end{equation}
Substituting (\rf{AmnExpand}) into (\rf{AmnEq}) and
 (\rf{anti self dual cond}), we get
\begin{equation}
g_1(r)=-2g_2'(r),\spc g_2(r)=-\frac{r^2-m^2}{4m}g_1(r),
\end{equation}
and the solution is
\begin{equation}
g_1(r)=c\frac{4m}{(r+m)^2},\spc
g_2(r)=-c\frac{r-m}{r+m}.
\end{equation}
We should determine the normalization factor $c$
by the condition that the charge of the string endpoint is one \cite{Sen}.
An open string attached on a D6-brane is
a membrane hooked on the center of the Kaluza-Klein monopole,
 where the $S^1$ fiber shrinks to a point%
\footnote{This ``hooking'' is related to an anomalous process
in which, when a D6 and a D2-brane pass through each other,
a string stretched between them
is created \cite{anomalous}.}.
For example, let us consider the following membrane configuration:
\begin{equation}
y^i=y^i(\sigma^0),\spc
r=\sigma^1,\spc
\psi=\sigma^2,\spc
\theta,\phi=\mbox{const}.
\end{equation}
This configuration represents a string attached on a D6-brane at
$y^i(\sigma^0)$
and stretched along the $r$-direction.
Using (\rf{bunri2}),
the contribution of this configuration to the action
 (\rf{SM2}) reads
\begin{equation}
S_{M2}=cg_{M2}\int drd\psi A_{r\psi}\int dy^iV_i
=4\pi cg_{M2}\int dy^iV_i.\spc
\end{equation}
Therefore, $c$ is fixed to be
\begin{equation}
c=\frac{1}{4\pi g_{M2}}=\frac{1}{2}\frac{1}{(2\pi)^3l_s^{\frac{3}{2}}}.
\end{equation}

Between these two solutions for $A_\mu$ and $A_{\mu\nu}$,
the following relation is holds:
\begin{equation}
\nabla_\mu A_\nu-\nabla_\nu A_\mu=-\frac{1}{2\pi l_s^2}A_{\mu\nu}.
\end{equation}
Using (\rf{bunri1}), (\rf{bunri2}) and this relation,
the 4-form field strength in (\rf{Maction})
is rewritten as
\begin{eqnarray}
K_{\mu\nu ij}
&=&A_{\mu\nu}(\nabla_iV_j-\nabla_jV_i)
  +(\nabla_\mu A_\nu-\nabla_\nu A_\mu)B_{ij}\nonumber\\
&=&A_{\mu\nu}V_{ij},
\lab{K4is}
\end{eqnarray}
and hence the action is
\begin{equation}
- -\frac{1}{8}\int\! d^{11}x K_{ij\mu\nu}K^{ij\mu\nu}
=-\frac{a}{8}\int\! d^7yV_{ij}^2,
\end{equation}
with
\begin{equation}
a=\int\! d^4x\,(A_{\mu\nu})^2=32\pi^2c^2=2\frac{1}{(2\pi)^4l_s^3}.
\end{equation}
Rescaling the metric, we get the desired action
which is identical to the second term in (\rf{BIexpand}):
\begin{equation}
S=-\frac{1}{4(2\pi)^4l_s^3e^{\phi_0}}\int\! d^7y V_{ij}^2.
\end{equation}

%%%%%%%%%%%%%%%%%%%%%%%%%%%%%%%%%%%%%%%%%%%%%%%%%%%%%%%%%%%%%%%%%%%%%%%%%%
\section{Chern-Simons terms}\lab{sec:CS}
In this section we shall show that the first three terms in the
Chern-Simons action (\rf{CSexpand}) are reproduced from M-theory.
The first term of (\rf{CSexpand}) is an ordinary magnetic coupling
of the R-R one form $C_M$.
The one form field $C_M$ is given by the eleven dimensional metric as
\begin{equation}
C_M=b\,\frac{g_{11,M}}{g_{11,11}},
\end{equation}
where $b$ is a normalization factor
and the electric charge is quantized in unit of $b$.
The kinetic term of $C_M$ is obtained from
the eleven dimensional Einstein action:
\begin{equation}
S=-\frac{b^2}{(2\pi)^7l_s^6}\frac{1}{4}F_{MN}^2.
\end{equation}
Comparing the coefficient in this equation to the one in (\rf{RRaction}),
the constant $b$ is determined as
\begin{equation}
b=(2\pi)^{\frac{7}{2}}l_s^3.
\end{equation}
Due to the Dirac condition, this means that the unit of magnetic charge is
 $2\pi/b=(2\pi)^{-\frac{5}{2}}l_s^{-3}$,
and the coupling corresponding to this charge is
\begin{equation}
S=\frac{1}{(2\pi)^{\frac{5}{2}}l_s^3}\frac{1}{2\cdot7!}
\int dy^{i_1}\wedge\cdots\wedge dy^{i_7}C_{i_1\cdots i_7}.
\lab{C7coupling}
\end{equation}
The agreement of (\rf{C7coupling}) with the first term of
(\rf{CSexpand}) is no wonder because the coefficient of this term is
determined only by the Dirac quantization condition.

%%%%%%%%%%%%%%%%%%%%%%%%%%%%%%%%%%%%%%%%%%%%%%%%%%

Next, let us consider the term
\begin{equation}
- -\frac{1}{2\cdot4!}\sqrt{-g}K_{KLMN}K^{KLMN}
=-\frac{1}{2\cdot4!\cdot7!}\epsilon^{ABCDEFGHIJK}K_{ABCD}\wt K_{EFGHIJK},
\end{equation}
in the eleven dimensional action (\rf{Maction}).
This term contains the part,
\begin{equation}
- -\frac{1}{2\cdot2\cdot5!}\epsilon^{\mu\nu ij\alpha abcde\beta}
K_{\mu\nu ij}\partial_\alpha A_{abcde\beta},
\lab{edAdA}
\end{equation}
where $K_{\mu\nu ij}$ is given by (\rf{K4is}),
and $A_{bcdef\mu}$ should be expressed as
a product of the vector zero mode $A_\mu$ of the last section
and the R-R 5-form $C_{abcde}$,
\begin{equation}
A_{abcde\mu}=kC_{abcde}A_\mu.
\lab{A6C5}
\end{equation}
The numerical factor $k$ is determined so that the kinetic term
of $C_{abcde}$ is
$-(2\cdot6!)^{-1}F_{abcdef}^2$ in the asymptotic region.
Substituting (\rf{A6C5}) into the kinetic term of $A_{bcdef\mu}$, we get
\begin{eqnarray}
- -\frac{1}{2\cdot7!}\int\! d^{11}x\sqrt{-g}\wt K_{ABCDEFG}\wt K^{ABCDEFG}
\!\!&=&\!\!
- -\frac{k^2}{2\cdot6!}\int\! dx^{11}\!\sqrt{g_{11,11}}\int\! d^{10}x
              \sqrt{-g}\,(F_{abcdef})^2(A_\mu)^2\nonumber\\
\!\!&=&\!\!-\frac{k^2}{2\cdot6!}
              \frac{e^{-\frac{2}{3}\phi_0}}{(2\pi)^7l_s^8}\int\! d^{10}x
              \sqrt{-g}\,(F_{abcdef})^2,
\lab{Ktilde^2}
\end{eqnarray}
where $F_{abcdef}=\partial_aC_{bcdef}+\mbox{$5$ terms}$.
Rescaling the metric, (\rf{Ktilde^2}) is reduced to
\begin{equation}
- -\frac{1}{2\cdot6!}
              \frac{k^2}{(2\pi)^7l_s^8}\int d^{10}x
              \sqrt{-g}(F_{abcdef})^2,
\end{equation}
and $k$ is determined to be
\begin{equation}
k=(2\pi)^{\frac{7}{2}}l_s^4.
\end{equation}
Then, substituting (\rf{K4is}) and (\rf{A6C5}) into
(\rf{edAdA}), we get
\begin{eqnarray}
S&=&\frac{(2\pi)^{\frac{7}{2}}l_s^4}{2}
    \int d^4x\epsilon^{\alpha\beta\mu\nu}
          A_{\mu\nu}\partial_\alpha A_\beta
   \times\frac{1}{2\cdot5!}\int d^7y
    \epsilon^{abcdeij}V_{ij}C_{bcdef}\nonumber\\
 &=&\frac{1}{(2\pi)^{\frac{3}{2}}l_s}
   \frac{1}{2\cdot5!}\int d^7y\epsilon^{abcdefg}C_{abcde}V_{fg}.
\end{eqnarray}
Thus we have reproduced the second term of (\rf{CSexpand}).

%%%%%%%%%%%%%%%%%%%%%%%%%%%%

The third term in (\rf{CSexpand}), $CVV$-term, can be extracted
from the eleven dimensional Chern-Simons term (\rf{MCS}).
First, we have to determine the rescaling factor between $A_{ijk}$ and
$C_{ijk}$. Comparing the kinetic terms of these fields in (\rf{Maction})
and (\rf{RRaction}), we get the relation:
\begin{equation}
A_{ijk}=\frac{1}{\sqrt{2\pi l_s}}C_{ijk}.
\lab{A3-C3}
\end{equation}
Because the three form field in seven dimension $A_{ijk}$
is a scalar on Taub-NUT,
its zero mode is a constant which does not depend on $x^\mu$.
Replacing the $A_{LMN}$ in (\rf{MCS}) with
(\rf{A3-C3}) and the two $K_{ABCD}$ with (\rf{K4is}),
we get exactly the third term of (\rf{CSexpand}):
\begin{eqnarray}
S&=&-\frac{\sqrt{2\kappa_{11}^2}}{\sqrt{2\pi l_s}}\times
  \frac{1}{4}\int d^4x
   \epsilon^{\mu\nu\rho\sigma}A_{\mu\nu}A_{\rho\sigma}\times
  \frac{1}{3!\cdot2\cdot2\cdot2}\int d^7y \epsilon^{ijklmnp}
  C_{ijk}V_{lm}V_{np}\nonumber\\
&=&\frac{l_s}{\sqrt{2\pi}}
  \frac{1}{3!\cdot2\cdot2\cdot2}\int d^7y \epsilon^{ijklmnp}
  C_{ijk}V_{lm}V_{np}.
\end{eqnarray}
%%%%%%%%%%%%%%%%%%%%%%%%%%%%%%%%%%%%%%%%%%%%%%%%%%%%%%%%%%%%%%%%%%%%%%%%%%%
\section{Discussion}
In this paper, we have shown that the $V^2$ term and the
$C+CV+CVV$ terms are correctly reproduced from the Kaluza-Klein
monopole.
We shall give some comments about the terms we have not
discussed in this paper.

The Born-Infeld action (\rf{BIexpand}) expanded to quadratic order
contains another term $(\partial\phi^a)^2$ where $\phi^a$ are scalar
fields on D6-brane and represent the transversal shifts of the brane.
These fields belong to the ${\bf 3}$ representation of the isometry
group $SU(2)$.
The shift of the center of the Kaluza-Klein monopole
implies that the metric varies.
Therefore, the $(L=1,N=0)$ mode of the metric would be relevant for
$\phi^a$.

Another term we have not discussed is the $CVVV$-term in
(\rf{CSexpand}).
This is cubic order in the vector field $V_i$,
which is a part of the three form field $A_{LMN}$ in eleven
dimension. The unique cubic term of the three form field 
in the action is (\rf{MCS}).
This term does not contain the metric and, apparently,
it seems impossible to extract the $CVVV$-term from this term.
However, the anti-self-duality condition (\rf{anti self dual cond})
of $A_{\mu\nu}$ depends on the metric
and the $AKK$-term can contain $C_\mu$ indirectly,
and hence the $CVVV$-term could be reproduced from the $AKK$-term.
To examine this possibility, we have to solve the zero mode
equations by taking account of the variation of the background
metric.

\vspace{2ex}
I would like to thank H.\ Hata
for valuable discussions and careful reading of the manuscript.

\vspace{2ex}
\noindent
{\bf Note Added:}
After this work was completed, the paper \cite{SigmaModel} appeared,
which also proposes the Born-Infeld effective action
of Kaluza-Klein monopole by a different approach from ours.

%%%%%%%%%%%%%%%%%%%%%%%%%%%%%%%%%%%%%%%%%%%%%%%%%%%%%%%%

%%%%%%%%%%%%%%%%%%%%%%%%%%%%%%%%%%%%%%%%%%%%%%%%%%%%%%%%%%%%%%%%%%%

\end{document}